\documentclass[conference]{IEEEtran}
\makeatletter
\def\ps@headings{%
\def\@oddhead{\mbox{}\scriptsize\rightmark \hfil \thepage}%
\def\@evenhead{\scriptsize\thepage \hfil \leftmark\mbox{}}%
\def\@oddfoot{}%
\def\@evenfoot{}}
\makeatother
\usepackage{cite}
\usepackage{citesort}
\usepackage{graphicx}
\usepackage{epstopdf}
\usepackage{amsmath}
\usepackage{xcolor}
\usepackage{amsfonts}
\usepackage{nicefrac}
\usepackage{amssymb,bm,upgreek}
\usepackage{algorithm}
\usepackage{flushend}
\usepackage{algpseudocode}

\begin{document}

\title{User Load Analysis and Pilot Sequence Design for Multi-Cell Massive MIMO Networks}

\author{\IEEEauthorblockN{Noman Akbar, Nan Yang, Parastoo Sadeghi, and Rodney A. Kennedy}
\IEEEauthorblockA{Research School of Engineering, Australian National University, Acton, ACT 2601,
Australia}Email: \{noman.akbar, nan.yang, parastoo.sadeghi, rodney.kennedy\}@anu.edu.au}

\markboth{Submitted to IEEE ICC 2016}{Akbar \MakeLowercase{\textit{et
al.}}: user load Analysis and Pilot Sequence Design for Multi-Cell Massive MIMO Networks}

\maketitle

\begin{abstract}
We propose a novel algorithm to design user load-achieving pilot sequences that mitigate pilot contamination in multi-cell massive multiple-input multiple-output (MIMO) networks. To this end, we first derive expressions for the user load and the load region of the network considering both small-scale and large-scale propagation effects. We then develop the pilot sequence algorithm for multi-cell massive MIMO networks as per the rules of generalized Welch bound equality design. Notably, we find that our algorithm and the corresponding downlink power allocation ensure that the user load is achieved when the signal-to-interference-plus-noise ratio (SINR) requirements for the users lie within the load region. Furthermore, we demonstrate the performance advantage of our proposed design relative to the existing designs, in terms of a larger load region and a higher maximum permitted SINR. Finally, we show that our proposed design can satisfy the pre-defined SINR requirements for users with a finite number of antennas at the base station (BS), while the existing designs cannot satisfy the same requirements even with an infinite number of antennas at the BS.
\end{abstract}

\IEEEpeerreviewmaketitle

\section{Introduction}\label{sec:intro}

Massive multiple-input multiple-output (MIMO) has emerged as one of the disruptive technologies for future fifth generation cellular networks, due to its potential benefits such as improvement of several orders of magnitude in spectral and energy efficiencies \cite{Boccardi2014}. The key idea behind massive MIMO is to deploy a very large number of antennas at the base station (BS) to serve many tens of users simultaneously. With such a deployment, massive MIMO reaps almost all the advantages offered by conventional MIMO, but on a much larger scale \cite{Yang2015}. A substantial implementation burden incurred by deploying hundreds of antennas is the channel estimation for a large number of channels. In order to ease this burden, the time division duplex (TDD) mode is adopted in massive MIMO networks such that the channels between the BS and the users are estimated via the uplink \cite{Lu2014}. Due to the assumption of channel reciprocity in the TDD mode, the estimated channel knowledge can be directly utilized for the downlink transmission \cite{Rusek2013}.

Pilot contamination is a practical problem that any massive MIMO network designer can face, which occurs when non-orthogonal pilot sequences are used across the whole network. In massive MIMO networks, the number of users is in general larger than the number of available pilot sequences. This indicates that the same pilot sequence needs to be assigned to two or more users, resulting in pilot contamination. Therefore, pilot contamination is identified as one of the main performance limiting factors in massive MIMO \cite{Marzetta2010,Larsson2014,Ashikhmin2012}. Several methods have been proposed to address the pilot contamination problem in massive MIMO networks,
such as protocol based methods \cite{Fernandes2013}, precoding based methods \cite{Jose2011}, angle-of-arrival based methods \cite{Yin2013}, and blind methods \cite{Muller2014}. While the aforementioned studies stand on their own merits, very little attention has been paid to the pilot sequence design, which decreases the negative effect of pilot contamination on the network performance. One example is \cite{Shen2015}, which designed pilot sequences and power allocation scheme for a single-cell massive MIMO network. For a multi-cell massive MIMO network, unfortunately, the design of load-achieving pilot sequences has not been explored in the literature. Despite its practical significance, such design is not trivial as multiple cells impose additional constraints that need to be satisfied, e.g., the per-cell quality of service requirements of users.

In this paper, we propose a novel pilot sequence design for an $L$-cell massive MIMO network. In each cell, an $N_{t}$-antenna BS communicates with $K$ single-antenna users in the TDD mode. In our design, the pilot sequences are generated for all the users in the network as per the rules of the generalized Welch bound equality (GWBE) sequence design \cite{Waldron2003}. The rationale behind choosing GWBE lies in its ability of achieving user capacity \cite{Ulukus2001} in code-division-multiple-access (CDMA) systems \cite{Cotae2006}. Notably, the GWBE design for a single cell CDMA system can not be directly utilized in multi-cell massive MIMO networks. Moreover, our design guarantees low correlation between different pilot sequences and thus reduces the detrimental impact of pilot contamination. The primary contributions of this paper are summarized as follows:
\begin{enumerate}
\item
We derive a new expression for the user load of the multi-cell massive MIMO network. The user load is defined as the number of users that can be simultaneously served, with their signal-to-interference-plus-noise ratio (SINR) requirements being satisfied. We then determine the load region of the network, under which the newly derived user load is achieved.
\item
We propose an easy-to-implement algorithm which produces load-achieving pilot sequences satisfying the SINR requirements at all the users within the network. Based on the algorithm, we also determine the power allocation for the downlink data transmission.
\item
We present numerical results to provide useful insights into the advantages of our proposed GWBE design over the existing pilot sequence designs. We show that our design achieves a larger load region and supports a greater range of SINR requirement than the existing designs. We further show that our design satisfies the SINR requirement with finite $N_{t}$, due to the larger load region it achieves, while the existing designs do not, even with infinite $N_{t}$.
\end{enumerate}

\section{Multi-Cell Massive MIMO Networks}\label{sec:system}

In this work we consider an $L$-cell TDD massive MIMO network. In each cell, an $N_{t}$-antenna BS communicates with $K$ single-antenna users. We denote $K_{tot}$ as the total number of users in the network, where $K_{tot}=KL$. In this network we consider both large-scale and small-scale propagation effects. Specifically, we denote $\sqrt{\beta_{i_{j}l}}h_{i_{j}l_{n}}$ as the propagation factor between the $j$th user in the $i$th cell and the $n$th BS antenna in the $l$th cell, where $i\in\left\{1,\dotsc,L\right\}$, $j\in\left\{1,\dotsc,K\right\}$, $l\in\left\{1,\dotsc,L\right\}$, and $n\in\left\{1,\dotsc,N_{t}\right\}$. Here, $\beta_{i_{j}l}$ characterizes the large-scale path loss effect from the $j$th user in the $i$th cell to the BS in the $l$th cell and $h_{i_{j}l_{n}}$ characterizes the small-scale multipath fading effect from the $j$th user in the $i$th cell to the $n$th BS antenna in the $l$th cell. In particular, we assume that $h_{i_{j}l_{n}}$ follows an independent and identically distributed (i.i.d) Rayleigh fading model, i.e., $h_{i_{j}l_{n}}\sim\mathcal{CN}(0,1)$. We also assume a block fading channel model, where the channel remains constant during the coherence time interval of $T$ but changes independently every interval. We further assume channel reciprocity between the uplink and the downlink, which is typical in TDD systems \cite{Jose2011}. Under this assumption, the propagation factor estimated via the uplink can be utilized for the downlink transmission.

\subsection{Channel Estimation via the Uplink}

We first focus on the channel estimation process in the uplink. In this process, the propagation factors in each cell are estimated by the BS using the pilot sequences sent by $K$ same-cell users. All pilot sequences are assumed to have unit energy and the length of $\tau$. We assume perfect synchronization between the uplink pilot sequences, which is regarded as the worst case scenario for pilot contamination  \cite{Yin2013}. Noticeably, synchronization errors result in decoration between the pilot sequences. The pilot sequence vector received at the $\textrm{BS}$ in the $l$th cell during the uplink training phase, denoted by a $\tau M \times 1$ vector, $\mathbf{s}_{l}$, is given by
\begin{align}\label{rec_pilot}
\mathbf{s}_{l}=\sum_{i=1}^{L}\sum_{j=1}^{K}\eta_{i_{j}l}{\mathbf{Q}_{i_{j}}}\mathbf{h}_{i_{j}l}+\mathbf{n}_{l},
\end{align}
where $\eta_{i_{j}l} = \sqrt{p_{i_{j}}\beta_{i_{j}l}}$, $\mathbf{Q}_{i_{j}} = {\mathbf{q}_{i_{j}}}\otimes \mathbf{I}_{N_{t}}$ is a $\tau N_{t} \times N_{t}$ matrix, $\mathbf{q}_{i_{j}}$ is the $\tau\times1$ pilot sequence assigned to the $j$th user in the $i$th cell, $\otimes$ denotes the Kronecker product, $\mathbf{I}_{N_{t}}$ denotes the $N_{t}\times{}N_{t}$ identity matrix, $p_{i_{j}}$ is the pilot power for the $j$th user in the $i$th cell, $\mathbf{h}_{i_{j}l}=[h_{i_{j}l_{1}},h_{i_{j}l_{2}},\dotsc,h_{i_{j}l_{n}}]^{T}$ is an $N_{t}\times1$ uplink channel vector from the $j$th user in the $i$th cell to the BS in the $l$th cell, and $\mathbf{n}_{l}$ is the $\tau{}N_{t}\times1$ additive white Gaussian noise (AWGN) at the BS in the $l$th cell.

We assume that the least square (LS) channel estimation method is adopted. It has been demonstrated that mean square error of an LS channel estimator remains nearly constant as $N_{t}$ increases \cite{Khansefid2015}, which makes it attractive for massive MIMO networks. Accordingly, the uplink channel from the $k$th user in the $l$th cell to the BS in the $l$th cell is obtained by utilizing the property of the pilot sequence matrix, given by $\mathbf{Q}_{l_{k}}^{T}\mathbf{Q}_{l_{k}}=\mathbf{I}_{N_{t}}$. Based on \eqref{rec_pilot} and assuming that the uplink power control is enabled with $\eta_{l_{k}l}=1$, we obtain the LS channel estimate as
\begin{align}\label{channel_estimate_2}
\mathbf{\hat{g}}_{l_{k}l}=\mathbf{Q}_{l_{k}}^{T}\mathbf{s}_{l}= \mathbf{h}_{l_{k}l}+\sum_{i,j \neq l,k}\eta_{i_{j}l}\rho_{i_{j}l_{k}}\mathbf{h}_{i_{j}l}+{\mathbf{Q}_{l_{k}}^T}\mathbf{n}_{l},
\end{align}
where $\mathbf{Q}_{l_{k}}^{T}$ denotes the transpose of $\mathbf{Q}_{l_{k}}$, ${\sum}_{i,j\neq{}l,k}=\sum_{i=1}^{L}\sum_{j=1}^{K}$ with $(i,j)\neq(l,k)$, and $\rho_{i_{j}l_{k}}$ is the correlation coefficient between pilot sequences, defined as $\rho_{i_{j}l_{k}}=\mathbf{q}_{l_{k}}^{T}\mathbf{q}_{i_{j}}$, $k \in \left\{1,2,\dotsc, K\right\}$. We note that the value of $\rho_{i_{j}l_{k}}$ varies from $+1$ to $-1$, where $+1$ and $-1$ indicate a perfect positive correlation and a perfect negative correlation between pilot sequences, respectively, while 0 indicates no correlation (or equivalently, orthogonal pilot sequences). It is evident from \eqref{channel_estimate_2} that the use of non-orthogonal pilot sequences for different users in the network, i.e., $\rho_{i_{j}l_{k}}\neq0$, contaminates the uplink channel estimate. This effect is referred to as pilot contamination, which significantly deteriorates the performance of massive MIMO networks.

\subsection{Data Transmission via the Downlink}

We now concentrate on the downlink data transmission. In this transmission, a data symbol $x_{l_{k}}$ is sent to the $k$th user in the $l$th cell from the same-cell BS. The transmit power of $x_{l_{k}}$ at the BS is given by $\mathbb{E}\left[x_{l_{k}}^H{x}_{l_{k}}\right] = P_{l_{k}}$, where $\mathbb{E}[\cdot]$ denotes expectation. We assume that the data symbols are uncorrelated zero mean symbols and linearly precoded by a precoding vector $\mathbf{a}$. Thus, the precoded downlink transmission received by the $k$th user in the $l$th cell is given by
\begin{align}\label{rec_initial}
\hat{y}_{l_{k}}=
\sum_{m=1}^{L}\sum_{n=1}^{K}\sqrt{\beta_{l_{k}m}}\mathbf{h}_{l_{k}m}^{H}\left(\mathbf{a}_{m_{n}}x_{m_{n}}\right)+w_{l_{k}},
\end{align}
where $w_{l_{k}}$ is the AWGN at the $k$th user in the $l$th cell. Assuming that only the statistical information of the channel is available at the user \cite{Jose2011,Shen2015}, we rewrite $\hat{y}_{l_{k}}$ in \eqref{rec_initial} as
\begin{align}\label{received_siga}
\hat{y}_{l_{k}}=\sqrt{\beta_{l_{k}l}}\mathbb{E}\left[\mathbf{h}_{l_{k}l}^H\mathbf{a}_{l_{k}}\right]x_{l_{k}}+u_{l_{k}},
\end{align}
where $u_{l_{k}}=\sqrt{\beta_{l_{k}l}}\left({\mathbf{h}_{l_{k}l}^H\mathbf{a}_{l_{k}}}
-\mathbb{E}\left[{\mathbf{h}_{l_{k}l}^H\mathbf{a}_{l_{k}}}\right]\right)x_{l_{k}}+\sum_{m,n \neq l,k}\sqrt{\beta_{l_{k}m}}\mathbf{h}_{l_{k}m}^{H}\left(\mathbf{a}_{m_{n}}x_{m_{n}}\right)+w_{l_{k}}$.
We clarify that the term $u_{l_{k}}$ can be treated as the effective noise and is uncorrelated with $\sqrt{\beta_{l_{k}l}}\mathbb{E}\left[{\mathbf{h}_{l_{k}l}}\right]x_{l_{k}}$. With the aid of \eqref{received_siga}, we determine the SINR at the users, evaluate the user load, and design the pilot sequences for the multi-cell massive MIMO network in Section \ref{sec:design}.

\section{User Load Analysis and Pilot Sequence Design}\label{sec:design}

In this section, we first derive a new expression for the user load in the multi-cell massive MIMO network. We then determine the load region of the network, under which the derived user load is achieved. We further propose an easy-to-implement algorithm to design pilot sequences that satisfy the SINR requirements at users and achieve the user load.


\subsection{Analysis of User Load}\label{sec:user_capacity_analysis}

Throughout this paper, the user load is defined as the number of users that can be simultaneously served via the downlink in the massive MIMO network such that SINR requirements of all the users are satisfied. Here, we preserve a practical assumption that the number of users in each cell is higher than the length of the pilot sequence, i.e., $K>\tau$. This is due to the fact that the massive MIMO BS typically serves a huge number of users using a limited number of pilot sequences. Under this assumption, the network performance suffers from both inter-cell pilot contamination and intra-cell pilot contamination, which is treated as a worst-case scenario for a pilot contaminated massive MIMO network.

\subsubsection{Signal-to-Interference-Plus-Noise Ratio at Users}

We commence our analysis by formulating the achievable SINR for the $k$th user in the $l$th cell, denoted by $\phi_{l_{k},N_{t}}$. Based on \eqref{received_siga}, we express $\phi_{l_{k},N_{t}}$ as
\begin{align}\label{long_exp}
\phi_{l_{k},N_{t}}=\frac{\left(\mathbb{E}\left[{\mathbf{h}_{l_{k}l}^{H}\mathbf{a}_{l_{k}}}\right]\right)^2\beta_{l_{k}l}P_{l_{k}}}
{\textrm{var}\left[{\mathbf{h}_{l_{k}l}^{H}\mathbf{a}_{l_{k}}}\right]\beta_{l_{k}l}P_{l_{k}} + \overline\phi_{l_{k},N_{t}} + \sigma_{w}^2},
\end{align}
where
\begin{align}\label{long_exp_phi_bar}
\hspace{-0.1cm}\overline\phi_{l_{k},N_{t}}= \sum_{m,n{}\neq{}l,k}\mathbb{E}\left[|{\mathbf{h}_{l_{k}m}^{H}\mathbf{a}_{m_{n}}}|^{2}\right]\beta_{l_{k}m}P_{m_{n}},
\end{align}
$\textrm{var}\left[\cdot\right]$ denotes the variance operation, and $\sigma_{w}^2$ is the variance of $w_{l_{k}}$.

We note that the achievable SINR given by \eqref{long_exp} is a generalized expression which is valid for any precoder. We now determine the achievable SINR with the maximum-ratio transmission
(MRT) precoder. Using \eqref{channel_estimate_2} and the channel hardening property of massive MIMO, the MRT precoding vector for the $k$th user in the $l$th cell is given by
\begin{align}\label{MRT_precoder}
\mathbf{a}_{l_{k}}=\frac{\mathbf{\hat{g}}_{l_{k}l}}{\|\mathbf{\hat{g}}_{l_{k}l}\|}=
\frac{\mathbf{\hat{g}}_{l_{k}l}}{\sqrt{N_{t}
\left(\mathbf{\hat{g}}_{l_{k}l}^{H}\mathbf{\hat{g}}_{l_{k}l}/N_{t}\right)}}
=\frac{\mathbf{\hat{g}}_{l_{k}l}}{\sqrt{N_{t}\delta_{l_{k}}}},
\end{align}
where $\|\cdot\|$ denotes the $l_2$ norm and $\delta_{l_{k}}=\sum_{i=1}^{L}\sum_{j=1}^{K}\eta_{i_{j}l}^2\rho_{i_{j}l_{k}}^{2}+\sigma_{n_{l}}^{2}$. Using the MRT precoding indicated by \eqref{MRT_precoder} together with the LS channel estimation, we present a simplified expression for $\phi_{l_{k},N_{t}}$ in the following Lemma.
\newtheorem{lemma}{Lemma}
\begin{lemma}\label{lemma_red}
If the MRT precoding is used with the LS channel estimation, the achievable SINR is simplified as
\begin{align}\label{SINR}
\phi_{l_{k},N_{t}}=\frac{\beta_{l_{k}l}P_{l_{k}}}{\delta_{l_{k}}\left[\sum\limits_{m,n\neq{}l,k} \frac{\rho_{l_{k}m_{n}}^{2}\eta_{l_{k}m}^2\beta_{l_{k}m}P_{m_{n}}}{\delta_{m_{n}}}+\frac{1}{N_{t}}\left(\overline{P}_{lk}\right)
\right]},
\end{align}
\end{lemma}
where $\overline{P}_{lk}=\sum_{m=1}^{L}\sum_{n=1}^{K}\beta_{l_{k}m}P_{m_{n}} + \sigma_{w}^2$.

We next present an asymptotic expression for the achievable SINR given in \eqref{SINR} when $N_{t}\rightarrow\infty$. We note that $N_{t}\rightarrow\infty$ is a valid and widely-adopted assumption in massive MIMO networks. Under this assumption, the asymptotic expression for $\phi_{l_{k},N_{t}}$, denoted by $\phi_{l_{k},\infty}$, is derived as
\begin{align}\label{SINR_infa}
\phi_{l_{k},\infty}=\frac{\beta_{l_{k}l}P_{l_{k}}}{\delta_{l_{k}}\left(\sum\limits_{m=1}^{L}
\sum\limits_{n=1}^{K}\frac{\rho_{l_{k}m_{n}}^{2}\eta_{l_{k}m}^2\beta_{l_{k}m}P_{m_{n}}}{\delta_{m_{n}}}\right)-\beta_{l_{k}l}P_{l_{k}}}.
\end{align}
The asymptotic expression given by \eqref{SINR_infa} reveals that the pilot contamination is a performance limiting factor in massive MIMO networks, since $\rho_{l_{k}m_{n}}$ still exists and deteriorates the performance even when $N_{t}\rightarrow\infty$.

\subsubsection{User Load of the Network}

We now analyze the user load. We first simplify $\phi_{l_{k},\infty}$ given by \eqref{SINR_infa} using uplink power control assumption and re-express it as
\begin{align}\label{SINR_infb}
\phi_{l_{k},\infty} \geq \overline{\phi}_{l_{k},\infty}=\frac{P_{l_{k}}}{\delta_{l_{k}}\textrm{tr}\left(\mathbf{q}_{l_{k}}^{T} \mathbf{Q}\mathbf{D}\mathbf{A}\mathbf{Q}^{T}\mathbf{q}_{l_{k}}\right)-P_{l_{k}}},
\end{align}
where $\textrm{tr}\left(\cdot\right)$ denotes the trace operation and $\mathbf{Q}$, $\mathbf{D}$, and $\mathbf{A}$ are block matrices given by $\mathbf{Q}=\left[\mathbf{Q}_{1},\dotsc, \mathbf{Q}_{l},\dotsc,\mathbf{Q}_{L}\right]$, $\mathbf{D}=\text{diag}\left[\mathbf{D}_{1},\dotsc, \mathbf{D}_{l},\dotsc,\mathbf{D}_{L}\right]$, and $\mathbf{A}=\text{diag}\left[\mathbf{A}_{1},\dotsc,\mathbf{A}_{l},\dotsc,\mathbf{A}_{L}\right]$, respectively. Here, $\mathbf{Q}_{l}$ is the pilot sequence matrix for the $K$ users in the $l$th cell, given by $\mathbf{Q}_{l}=\left[\mathbf{q}_{l_{1}},\mathbf{q}_{l_{2}},\dotsc,\mathbf{q}_{l_{K}}\right]$, $\mathbf{D}_{l}$ is a diagonal matrix consisting of the transmit power at the BS in the $l$th cell for the $K$ same-cell users, given by $\mathbf{D}_{l}=\textrm{diag}\left\{P_{l_{1}}, P_{l_{2}}, \dotsc, P_{l_{K}}\right\}$, and $\mathbf{A}_{l}$ is a diagonal matrix consisting the inverse of parameter $\delta_{l_{k}}$ for all the $K$ users in the $l$th cell, given by $\mathbf{A}_{l}=\textrm{diag}\left\{1/\delta_{l_{1}}, 1/\delta_{l_{2}}, \dotsc, 1/\delta_{l_{K}}\right\}$, where $\textrm{diag}\{\cdot\}$ denotes a diagonal matrix with indicated elements along the diagonal.

Based on \eqref{SINR_infb}, we find that
\begin{align}\label{interme-appendix6}
\hspace{-0.2cm}\sum_{i=1}^{L}\sum_{j=1}^{K}\left(\frac{1+\overline{\phi}_{i_{j},\infty}}{\overline{\phi}_{i_{j},\infty}}\right)
=\textrm{tr}\left(\mathbf{D}^{-1}\mathbf{A}^{-1}\mathbf{Q}^{T}\mathbf{Q} \mathbf{D}\mathbf{A}\mathbf{Q}^{T}\mathbf{Q}\right).
\end{align}
By defining $\mathbf{R}_{S}\triangleq\mathbf{Q}^{T}\mathbf{Q}$ and $\mathbf{Z}\triangleq\mathbf{D}\mathbf{A}$ in \eqref{interme-appendix6}, we obtain
\begin{align}\label{interme-appendix7}
\sum_{i=1}^{L}\sum_{j=1}^{K}\left(\frac{1+\overline{\phi}_{i_{j},\infty}}{\overline{\phi}_{i_{j},\infty}}\right) &=\textrm{tr}\left(\mathbf{Z}^{-1}\mathbf{R}_{S}\mathbf{Z}\mathbf{R}_{S}\right),\notag\\
&\geq{}K_{tot}+\underbrace{\sum_{p=1}^{L}\sum_{q=1}^{K}\sum_{r=1}^{L}\sum_{s=1}^{K}}_{p>r,q>s}2\rho_{p_{q}r_{s}}^2,\notag\\
&\geq\textrm{tr}\left(\mathbf{R}_{S}\mathbf{R}_{S}\right).
\end{align}
We note that $\mathbf{R}_{S}$ in \eqref{interme-appendix7} is a symmetric matrix. By performing the eigen-decomposition of $\mathbf{R}_{S}$, we simplify \eqref{interme-appendix7} as
\begin{align}\label{interme-appendix8}
\sum_{i=1}^{L}\sum_{j=1}^{K}\left(\frac{1+\overline{\phi}_{i_{j},\infty}}{\overline{\phi}_{i_{j},\infty}}\right)\geq \sum_{i=1}^{K_{tot}}\lambda_{i}^{2}=\frac{1}{\tau}K_{tot}^{2},
\end{align}
where $\lambda_{i}$ is the $i$th eigenvalue of $\mathbf{R}_{S}$.

We denote $\gamma_{i_{j}}$ as the SINR requirement for the $j$th user in the $i$th cell. As such, the achievable SINR with infinite $N_{t}$ needs to be higher than or equal to $\gamma_{i_{j}}$, i.e., $\phi_{i_{j},\infty} \geq \overline{\phi}_{i_{j},\infty}\geq\gamma_{i_{j}}$. This indicates that $\sum_{i=1}^{L}\sum_{j=1}^{K}\left(\frac{1+\gamma_{i_{j}}}{\gamma_{i_{j}}}\right)
\geq\sum_{i=1}^{L}\sum_{j=1}^{K}\left(\frac{1+\overline{\phi}_{i_{j},\infty}}{\overline{\phi}_{i_{j},\infty}}\right)$. Using this inequality, \eqref{interme-appendix8} can be rewritten as
\begin{align}\label{interme-appendix10}
K_{tot}\leq\sqrt{\tau\sum_{i=1}^{L}\sum_{j=1}^{K}\left(\frac{1+\gamma_{i_{j}}}{\gamma_{i_{j}}}\right)}.
\end{align}
We clarify that \eqref{interme-appendix10} gives an upper bound on the user load of a multi-cell massive MIMO network.

\subsubsection{Load Region of the Network}

We next determine the load region of the network under which the previously derived user load can be achieved. By applying the Cauchy-Schwarz inequality, it is proven that the user load indicated by \eqref{interme-appendix10} is always achieved when the following condition holds:
\begin{align}\label{BW_all}
\sum_{i=1}^{L}\sum_{j=1}^{K}\left(\frac{\gamma_{i_{j}}}{1+\gamma_{i_{j}}}\right)\leq\tau.
\end{align}
We refer to the bound given by \eqref{BW_all} as the load region of the network. Under the load region, the user load given by \eqref{interme-appendix10} is achieved. Assuming that the load region is equally divided among the $L$ cells in the network, the upper bound on the load region for the $i$th cell is given by
\begin{align}\label{BW}
\sum_{j=1}^{K}\left(\frac{\gamma_{i_{j}}}{1+\gamma_{i_{j}}}\right)\leq\frac{\tau}{L}.
\end{align}

\subsection{Design of Pilot Sequences}\label{sec:pilot_sequence_design}

In this subsection, we propose an easy-to-implement algorithm to design the load-achieving pilot sequences for the multi-cell massive MIMO network. Here, we define the load-achieving pilot sequences as the sequences that satisfy the SINR requirements for all the users in the network and achieve the user load given by \eqref{interme-appendix10}.

In order to design the load-achieving pilot sequences, we define two $1\times{}K$ vectors $\mathbf{z}$ and $\mathbf{x}$, given by $\mathbf{z}=\left[\gamma_{l_{1}}/\left(1+\gamma_{l_{1}}\right),\gamma_{l_{2}}/\left(1+\gamma_{l_{2}}\right),\dotsc,\gamma_{l_{K}}/ \left(1+\gamma_{l_{K}}\right) \right]$ and $\mathbf{x}=\left[x_{1},x_{2},\cdots,x_{\tau},0,\cdots,0 \right]$, where $\gamma_{l_{1}}\geq\gamma_{l_{2}}\geq\dotsc\geq\gamma_{l_{K}}$ and $x_1\geq x_2\geq\dotsc\geq x_{\tau}$. We highlight that the values for $\gamma_{l_{k}}$ need to be chosen to satisfy \eqref{BW} with equality. We next present three preliminaries based on $\mathbf{x}$ and $\mathbf{z}$, as follows:
\newtheorem{preliminary}{Preliminary}
\begin{preliminary}\label{prem1}
Given vectors $\mathbf{z}$ and $\mathbf{x}$, $\mathbf{x}$ majorizes $\mathbf{z}$, i.e., $\mathbf{x}\succ\mathbf{z}$, if $\sum_{n=1}^{m}x_{n}\geq\sum_{n=1}^{m}z_{n}$, where $m\in\left\{1,\cdots,K-1\right\}$ and $\sum_{n=1}^{m}x_{n}=\sum_{n=1}^{m}z_{n}$, where $m=K$.
\end{preliminary}
\begin{preliminary}\label{prem2}
Given a vector $\mathbf{z}$, a vector $\mathbf{x}$ can be found for the value of $m=\tau$ such that  $\mathbf{x}\succ\mathbf{z}$, if the vector $\mathbf{x}$ is given by $x_{i}= \sum_{n=1}^{K}\left(z_{n}/\tau\right)$, where $i\in\left\{1,\cdots,\tau\right\}$ and $x_{j}=0$, where $j\in\left\{\tau+1,\cdots,K\right\}$.
\end{preliminary}
\begin{preliminary}\label{prem3}
If $\mathbf{x}\succ\mathbf{z}$, $\mathbf{z}$ is obtainable by applying at most $K-1$ T--transform operation on $\mathbf{x}$, i.e., $\mathbf{z}=\mathbf{T}_{K-1}\mathbf{T}_{K-2}\cdots\mathbf{T}_{1}\mathbf{x}$, and there exists a matrix $\mathbf{U}=\mathbf{U}_{1}\mathbf{U}_{2}\cdots\mathbf{U}_{K-1}$, where $\mathbf{U}_{i}$ is a unitary matrix generated from $\mathbf{T}_{i}$ at each step of the T-transform \cite{Viswanath1999}.
\end{preliminary}

We now present the step-by-step procedure of the load-achieving pilot sequences design for the multi-cell massive MIMO network. Specifically, this procedure is detailed in \textbf{Algorithm~\ref{algo1}}. This algorithm uses a $1\times K_{tot}$ vector $\mathbf{\Gamma}=\left[\pmb{\gamma}_{1},\cdots,\pmb{\gamma}_{l},\cdots,\pmb{\gamma}_{L}\right]$ and the length of the pilot sequence $\tau$ as the input, where  $\pmb{\gamma}_{l}=\left[\gamma_{l_1},\gamma_{l_2},\cdots,\gamma_{l_K}\right]$ is the $1\times{}K$ vector containing the minimum SINR requirements for the $K$ users in the $l$th cell. We clarify that the values of $\gamma_{l_k}$ need to be chosen such that \eqref{BW} is satisfied with equality, which in turn guarantees that \eqref{interme-appendix10} is achieved. The algorithm returns the pilot sequence matrix for the network, $\mathbf{Q}$, as the output.
\begin{algorithm}
\caption{Load--achieving pilot sequence design} \label{algo1}
\begin{algorithmic}[1]
\Procedure{Pilot Design}{$\mathbf{\Gamma},\tau$}\Comment{$\mathbf{\Gamma} = \left[\pmb{\gamma}_{1},\cdots,\pmb{\gamma}_{L}\right]$}
\For{$l\gets 1, L$}
\State $\pmb{\gamma} \gets  \pmb{\gamma}_{l}$
\Comment{where $\gamma_{ij}\leq \nicefrac{1}{L-1}$}
\State $\textrm{sum} \gets 0$
\State $\mathbf{x}_{l} \gets \mathbf{0}_{1\times K}$ \Comment{$\mathbf{x}_{l}$ is a $1 \times K$ zero vector}
\State $\mathbf{z}_{l} \gets \mathbf{0}_{1\times K}$ \Comment{$\mathbf{z}_{l}$ is a $1 \times K$ zero vector}
\For{$k\gets 1, K$}
\State $\mathbf{z}_{l}(k)\gets \left(\nicefrac{\gamma_{k}}{1+\gamma_{k}}\right)$ 
\State $\textrm{sum} \gets \textrm{sum} + \mathbf{z}_{l}(k)$
\EndFor
\State $B_{l} \gets \frac{\textrm{sum}}{\tau}$
\State $\mathbf{x}_{l}\left(1,\cdots,\tau\right) \gets B_{l}$ \Comment{\parbox[t]{.40\linewidth}{First $\tau$ elements of $\mathbf{x}_{l}$ are set to $B_{l}$}}
\State $\mathbf{U}_{l} \gets \Call{T-transform}{\mathbf{z}_{l},\mathbf{x}_{l}}$ 
\State $\mathbf{V}_{l} \gets \mathbf{U}_{l}\left(\tau,:\right)$ \Comment{\parbox[t]{.48\linewidth}{$\mathbf{V}_{l}$ retains first $\tau$ rows of $\mathbf{U}_{l}$}}
\State $\mathbf{Z}_{l} \gets \text{diag}\{\mathbf{z}_{l}\}$ \Comment{$\mathbf{z}_{l}$ is a diagonal matrix}
\State $\mathbf{Q}_{l} \gets \text{normc}\left(B_{l}^{\frac{1}{2}}\mathbf{V}_{l} \mathbf{Z}_{l}^{-\frac{1}{2}}\right)$
\EndFor
\State $\mathbf{Q} \gets \left[\mathbf{Q}_{1},\cdots,\mathbf{Q}_{L}\right]$ \Comment{\parbox[t]{.45\linewidth}{$\mathbf{Q}$ is the desired pilot sequence matrix}}
\State \textbf{return} $\mathbf{Q}$
\EndProcedure
\end{algorithmic}
\end{algorithm}

\textbf{Algorithm~\ref{algo1}} considers the SINR requirements in one cell at each time. First, the algorithm obtains the SINR requirements for all the $K$ users in the $l$th cell. Second, the algorithm calculates the effective bandwidth for all the $K$ users in the $l$th cell and sets the values in $\mathbf{z}_{l}$. Third, the algorithm finds $\mathbf{x}_{l}$ from $\mathbf{z}_{l}$ using \emph{Preliminary \ref{prem2}}. A $K\times K$ matrix $\mathbf{U}_{l}$ is found by applying the T-transform to $\mathbf{x}_{l}$ and $\mathbf{z}_{l}$, as described in \emph{Preliminary \ref{prem3}}. The $\tau \times K$ matrix $\mathbf{V}_{l}$ only retains the first $\tau$ rows of the vector $\mathbf{U}_{l}$. Finally, the pilot sequence matrix for all the $K$ users in the $l$th cell is found by normalizing the columns of the matrix given by $B_{l}^{\frac{1}{2}}\mathbf{V}_{l}\mathbf{Z}_{l}^{-\frac{1}{2}}$, i.e., $\text{normc}\left(B_{l}^{\frac{1}{2}}\mathbf{V}_{l} \mathbf{Z}_{l}^{-\frac{1}{2}}\right)$, such that each column has unit energy.
The process is repeated for all the $L$ cells until $\mathbf{Q}$ is obtained for the whole network. The finally obtained pilot sequences are known as the GWBE pilot sequences, which are capable of achieving the user load. Notably, \textbf{Algorithm~\ref{algo1}} can be implemented among $L$ BSs in a distributed manner. In the distributed implementation, the BS in the $i$th cell can design $\mathbf{Q}_{i}$ for all the $K$ users in the $i$th cell, without requiring any feedback from other BSs.

We note that \textbf{Algorithm~\ref{algo1}} designs pilot sequences for the case that \eqref{BW} is satisfied with equality. We now focus on the case where \eqref{BW} is not satisfied with equality. Since $\gamma_{i_{j}}/\left(1+\gamma_{i_{j}}\right)$ is a monotonically increasing function, there exists some value, $\hat{\gamma}_{i_{j}}\geq\gamma_{i_{j}}$, such that the equality in \eqref{BW} holds, i.e., $\sum_{j=1}^{K}\left(\frac{\hat\gamma_{i_{j}}}{1+\hat\gamma_{i_{j}}}\right)=\frac{\tau}{L}$. The value of $\hat{\gamma}_{i_{j}}$ is then used in our proposed GWBE design. Accordingly, the downlink transmit power for the $j$th user in the $i$th cell is set as $P_{i_{j}}=\delta_{i_{j}}\hat\gamma_{i_{j}}/\left(1+\hat\gamma_{i_{j}}\right)$.
We clarify that the use of $\hat{\gamma}_{i_{j}}$ also guarantees that the SINR requirements for all the users in the network are met and the user load of the network is achieved.


\section{Numerical Results}\label{sec:Numerical}

In this section we present numerical results that demonstrate the advantage of our proposed pilot sequence design over the current designs. Specifically, we compare the performance of the proposed GWBE design with the performance of two well-known pilot sequence designs, namely, the Welch bound equality (WBE) design and the finite orthogonal set (FOS) design. In the WBE design \cite{Sarwate1999}, the correlation coefficient between different pilot sequences is fixed, which is given by $\rho_{i_{j}l_{k}}=\sqrt{\left(K-\tau\right)/\left(\left(K-1\right)\tau\right)}$, where $\left(i,j\right)\neq\left(l,k \right)$. Therefore, the value of $\delta_{i_{j}}$ is the same for all the users in the network, i.e., $\delta_{i_{j}}=\delta$. In the FOS design \cite{Hoydis2011}, the correlation coefficient between different pilot sequences is always zero. Hence, only the users with the same pilot sequence are considered in the comparison presented in this section. We note that the load regions of the WBE design and the FOS design in a single-cell massive MIMO network have already been found in \cite{Shen2015}. Here, we extend the results in \cite{Shen2015} to a multi-cell network, which facilitate our comparison. Throughout this section, we consider that $K=4$ users in each cell and the length of the pilot sequence is $\tau=3$.

\begin{figure}[!t]
\centering
\includegraphics[height=2.5in,width=3.2in]{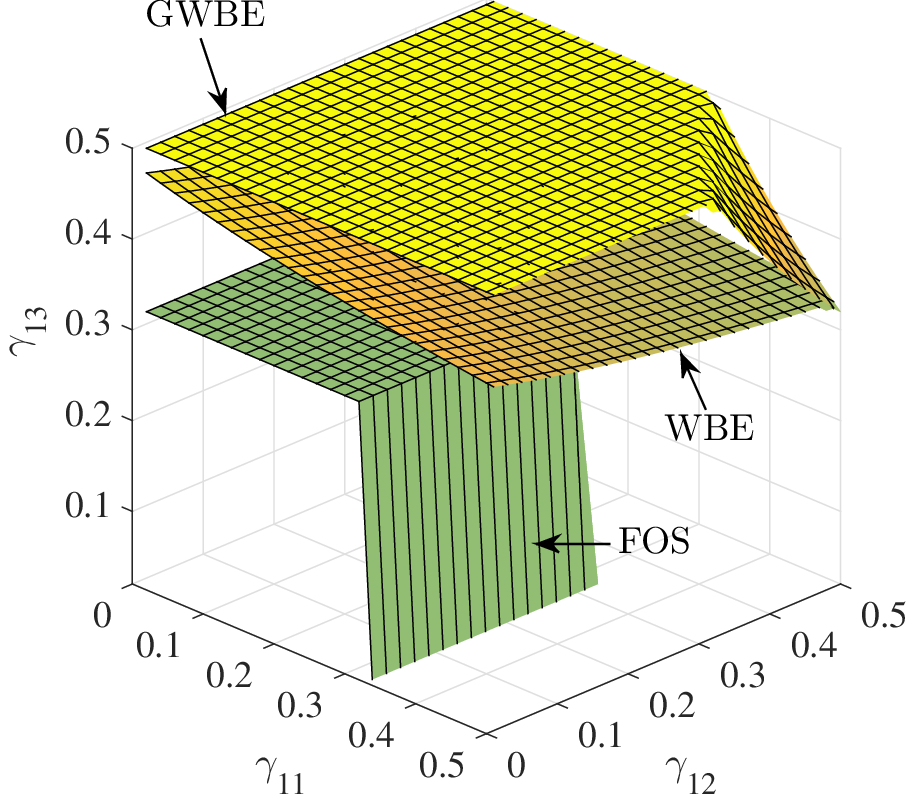}
\caption{The upper surface boundary of the load regions versus the SINR requirements for the proposed GWBE design and the existing WBE and FOS designs.}
\label{capacity_region}
\end{figure}

We first compare the load region of the proposed GWBE design with those of the WBE and FOS designs. In this comparison we consider a multi-cell massive MIMO network with $L=3$ cells such that $K_{tot}=12$. The SINR requirements for the users in 3 cells are set to $\pmb{\gamma}_{1}=\pmb{\gamma}_{2}=\pmb{\gamma}_{3}=\left[\gamma_{1_{1}},\gamma_{1_{2}},\gamma_{1_{3}},0.1\right]$. For our GWBE design, the values of $\gamma_{i_{j}}$ need to be
carefully chosen such that the majorization condition given by \emph{Preliminary \ref{prem1}} always holds, i.e., $\gamma_{i_{j}}\leq{1}/\left(L-1\right)$. For the FOS design, it is assumed that one pilot sequence is simultaneously used by 6 users in the network. Fig.~\ref{capacity_region} depicts the upper surface boundary of the load regions for three designs. We find that the load region for the proposed GWBE design is $20.9\%$ and $73.5\%$ larger than the WBE and FOS designs, respectively. Importantly, a larger load region indicates that a group of users with higher SINR requirements can be simultaneously served in a pilot contaminated massive MIMO network. This is due to the fact that a larger load region offers more freedom for the users to choose their required SINR level in the network without derogating from the limitation imposed by pilot contamination.

\begin{figure}[!t]
\centering
\includegraphics[height=2.5in,width=3.2in]{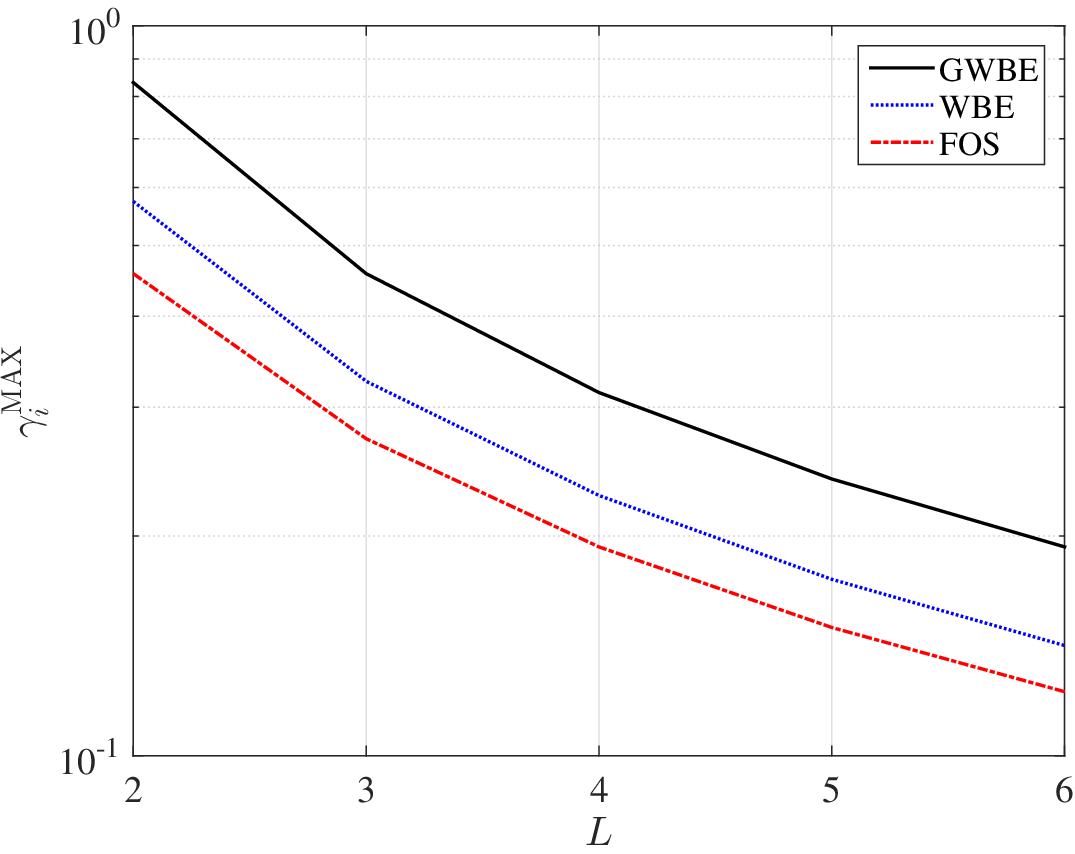}
\caption{The maximum permitted SINR versus the number of cells for the proposed GWBE design and the existing WBE and FOS designs.}
\label{num_cells}
\end{figure}

We now examine the effect of increasing the number of cells on the maximum permitted SINR in the $i$th cell, denoted by $\gamma_{i}^{\textrm{MAX}}$. Here, $\gamma_{i}^{\text{MAX}}$ is defined as the maximum value of the SINR requirements among the $K$ users in the $i$th cell, i.e. $\gamma_{i}^{\textrm{MAX}}=\max_{1\leq{j}\leq{K}}\gamma_{i_{j}}$.
In this examination, the SINR requirements for the users in the $L$ cells are given by $\pmb{\gamma}_{i}=\left[\gamma,\gamma,\gamma/2,\gamma/2\right]$, $\forall i\in\left\{1,\dots,L\right\}$. As such, we have $\gamma_{i}^{\textrm{MAX}}=\gamma$. Fig.~\ref{num_cells} depicts the maximum permitted SINR when the number of cells increases. 
We first observe that the maximum permitted SINR decreases when $L$ increases. For example, when $L$ increases from 2 to 6, the maximum permitted SINR of our GWBE design decreases from 0.84 to 0.19. This is due to the fact that increasing $L$ reduces the load region, which in turn restricts the maximum permitted SINR. Second, we observe that the proposed GWBE design achieves a higher maximum permitted SINR than the WBE and FOS designs. For example, when $L=2$, the maximum permitted SINR of our GWBE design is 0.84, while those of the WBE and FOS designs are 0.57 and 0.46, respectively. Crucially, a higher maximum permitted SINR indicates that the network can support higher SINR requirements. 

\begin{figure}[!t]
\centering
\includegraphics[height=2.5in,width=3.2in]{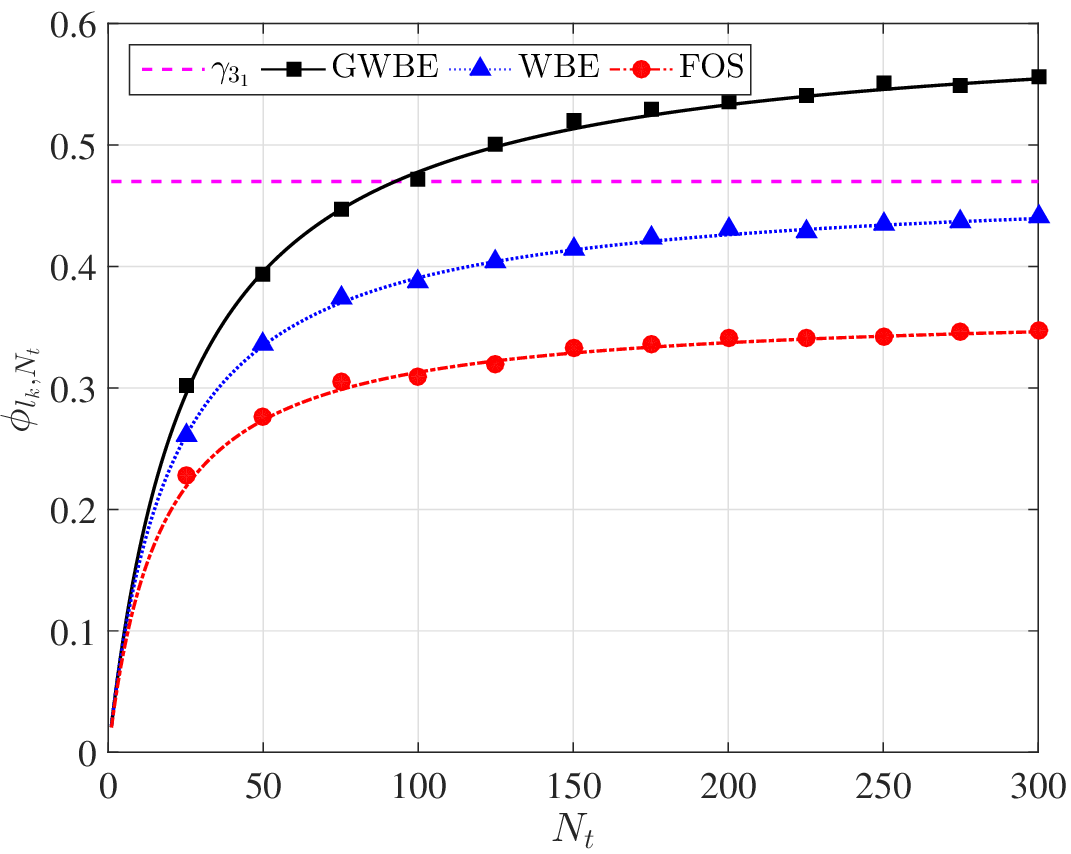}
\caption{The achievable SINR versus the number of antennas for the first user in the third cell for the proposed GWBE design and the existing WBE and FOS designs.}
\label{achievable SINR}
\end{figure}

Finally, we compare the achievable SINR with a finite number of antennas of the proposed GWBE design with those of the WBE and FOS designs. In this comparison we consider $L=3$, $\sigma_{w}^{2}=p_{l_{k}}=1$, $\beta_{l_{k}m}=1$, where $l=m$, and $\beta_{l_{k}m}=0.9$, where $l\neq m$. The SINR requirements for the users in 3 cells are set to $\pmb{\gamma}_{1}=\left[0.45, 0.38, 0.25, 0.19\right]$, $\pmb{\gamma}_{2}=\left[0.43, 0.38, 0.28, 0.20\right]$, and $\pmb{\gamma}_{3}=\left[0.47, 0.43, 0.28, 0.13\right]$. We clarify that these SINR requirements remain within the load region for the proposed GWBE design but lie outside the load region for the WBE and FOS pilot designs. This is not surprising since the load region of our GWBE design is larger than those of the WBE and FOS designs, as depicted in Fig.~\ref{capacity_region}. As such, the SINR requirements considered in this comparison demonstrate how a diverse range of SINR requirements can be satisfied by our GWBE design.
As stated in Section~\ref{sec:pilot_sequence_design}, the load-achieving pilot design procedure of our design requires that the value of $\hat{\gamma}_{l_{k}}>\gamma_{l_{k}}$ needs to be chosen such that \eqref{BW_all} is satisfied with equality. As such, we choose $\hat{\pmb{\gamma}}$ as $\pmb{\hat{\gamma}}_{1}=\left[0.48, 0.40, 0.27, 0.21\right]$, $\pmb{\hat{\gamma}}_{2}=\left[0.45, 0.40, 0.30, 0.22\right]$, and
$\pmb{\hat{\gamma}}_{3}=\left[0.49, 0.45, 0.30, 0.15\right]$.
Fig.~\ref{achievable SINR} depicts the achievable SINR for the first user in the third cell using the three designs. A key observation from Fig.~\ref{achievable SINR} is that only the proposed GWBE design satisfies the SINR requirement for the user when the number of antennas exceeds some threshold, i.e., $\phi_{3_{1}}>\gamma_{3_{1}}$ when $N_{t}\geq 93$. In contrast, the WBE and FOS designs do not satisfy the SINR requirement for the user, no matter how large $N_{t}$ is. Notably, the achievable SINR for the WBE and FOS designs using an infinite number of antennas is still below the SINR requirement.

\section{Conclusions}\label{sec:con}

We proposed a novel GWBE design that generates low correlated pilot sequences to address the pilot contamination problem in multi-cell massive MIMO networks. We first derived a new expression for the user load, determined the load region of the network, and then designed an algorithm to produce load-achieving pilot sequences satisfying the SINR requirements for the users. Using numerical results we demonstrated the advantage of our proposed GWBE design over the existing WBE and FOS designs. In particular, we showed that our GWBE design achieves a higher load region and supports a greater range of SINR requirements than the existing designs. We further showed that our GWBE design can satisfy a higher pre-determined SINR requirement with a finite number of antennas, while the other two designs cannot satisfy the same requirements even with an infinite number of antennas.

\end{document}